\input harvmac
%\draftmode
\def\journal#1&#2(#3){\unskip, \sl #1\ \bf #2 \rm(19#3) }
\def\andjournal#1&#2(#3){\sl #1~\bf #2 \rm (19#3) }

\def\ie{{\it i.e.}}
\def\eg{{\it e.g.}}

\def\frac#1#2{{#1\over#2}}

\def\half{\frac12}

\def\inbar{\,\vrule height1.5ex width.4pt depth0pt}
\def\IC{\relax\hbox{$\inbar\kern-.3em{\rm C}$}}
\def\IR{\relax{\rm I\kern-.18em R}}
\def\IP{\relax{\rm I\kern-.18em P}}
\def\IZ{\relax{\rm I\kern-.18em Z}}
\def\Z{{\bf Z}}

%
%%%%%%%%%%%%%%%%%%%%%%%%%%%%%%%%%%%%
%

%
\catcode`\@=11
\def\slash#1{\mathord{\mathpalette\c@ncel{#1}}}
\overfullrule=0pt

\def\LL{{\cal L}}

\def\SS{{\cal S}}

\def\underrel#1\over#2{\mathrel{\mathop{\kern\z@#1}\limits_{#2}}}

\catcode`\@=12

%%%%%%%%%%%%%%%%%%%%%%%%%%%%%%%%%%%%%%%%%%%%%%%%%%%%%%%%%%%%%%

%

\def\det{{\rm det}}

\def \cosh{{\rm cosh}}

\def\det{{\rm det}}
\def\exp{{\rm exp}}

%%%%%%%%%%%%%%%%%%%%%%%%%%%%%%%%%%%%%%%%%%%%%%%%%%%%%%%%%%%%%%
% new defs:

%\KutasovER
\lref\KutasovER{
D.~Kutasov and V.~Niarchos,
``Tachyon effective actions in open string theory,''
Nucl.\ Phys.\ B {\bf 666}, 56 (2003)
[arXiv:hep-th/0304045].
%%CITATION = HEP-TH 0304045;%%
}

%\NiarchosRW
\lref\NiarchosRW{
V.~Niarchos,
``Notes on tachyon effective actions and Veneziano amplitudes,''
hep-th/0401066.
%%CITATION = HEP-TH 0401066;%%
}

%\LukyanovNJ
\lref\LukyanovNJ{
S.~L.~Lukyanov, E.~S.~Vitchev and A.~B.~Zamolodchikov,
``Integrable model of boundary interaction: The paperclip,''
arXiv:hep-th/0312168.
%%CITATION = HEP-TH 0312168;%%
}

%\CallanAT
\lref\CallanAT{
C.~G.~.~Callan, J.~A.~Harvey and A.~Strominger,
``Supersymmetric string solitons,''
arXiv:hep-th/9112030.
%%CITATION = HEP-TH 9112030;%%
}

%\SenAN
\lref\SenAN{
A.~Sen,
``Field theory of tachyon matter,''
Mod.\ Phys.\ Lett.\ A {\bf 17}, 1797 (2002)
[arXiv:hep-th/0204143].
%%CITATION = HEP-TH 0204143;%%
}

%\PolchinskiRQ
\lref\PolchinskiRQ{
J.~Polchinski,
``String Theory. Vol. 1: An Introduction To The Bosonic String,''
Cambridge University Press, 1998.
%\href{http://www.slac.stanford.edu/spires/find/hep/www?irn=4634799}{SPIRES entry}
}

%\PolchinskiRR
\lref\PolchinskiRR{
J.~Polchinski,
``String Theory. Vol. 2: Superstring Theory And Beyond,''
Cambridge University Press, 1998.
%\href{http://www.slac.stanford.edu/spires/find/hep/www?irn=4634802}{SPIRES entry}
}

%\TseytlinDJ
\lref\TseytlinDJ{
A.~A.~Tseytlin,
``Born-Infeld action, supersymmetry and string theory,''
arXiv:hep-th/9908105.
%%CITATION = HEP-TH 9908105;%%
}

%\FelderSV
\lref\FelderSV{
G.~N.~Felder, L.~Kofman and A.~Starobinsky,
``Caustics in tachyon matter and other Born-Infeld scalars,''
JHEP {\bf 0209}, 026 (2002)
[arXiv:hep-th/0208019].
%%CITATION = HEP-TH 0208019;%%
}

%\FelderXU
\lref\FelderXU{
G.~N.~Felder and L.~Kofman,
``Inhomogeneous fragmentation of the rolling tachyon,''
arXiv:hep-th/0403073.
%%CITATION = HEP-TH 0403073;%%
}

%\SenXS
\lref\SenXS{
A.~Sen,
``Open-closed duality at tree level,''
Phys.\ Rev.\ Lett.\  {\bf 91}, 181601 (2003)
[arXiv:hep-th/0306137].
%%CITATION = HEP-TH 0306137;%%
}

%\SenIV
\lref\SenIV{
A.~Sen,
``Open-closed duality: Lessons from matrix model,''
Mod.\ Phys.\ Lett.\ A {\bf 19}, 841 (2004)
[arXiv:hep-th/0308068].
%%CITATION = HEP-TH 0308068;%%
}

%\SenNU
\lref\SenNU{
A.~Sen,
``Rolling tachyon,''
JHEP {\bf 0204}, 048 (2002)
[arXiv:hep-th/0203211].
%%CITATION = HEP-TH 0203211;%%
}

%\SenIN
\lref\SenIN{
A.~Sen,
``Tachyon matter,''
JHEP {\bf 0207}, 065 (2002)
[arXiv:hep-th/0203265].
%%CITATION = HEP-TH 0203265;%%
}

%\LarsenWC
\lref\LarsenWC{
F.~Larsen, A.~Naqvi and S.~Terashima,
``Rolling tachyons and decaying branes,''
JHEP {\bf 0302}, 039 (2003)
[arXiv:hep-th/0212248].
%%CITATION = HEP-TH 0212248;%%
}

%\OkudaYD
\lref\OkudaYD{
T.~Okuda and S.~Sugimoto,
``Coupling of rolling tachyon to closed strings,''
Nucl.\ Phys.\ B {\bf 647}, 101 (2002)
[arXiv:hep-th/0208196].
%%CITATION = HEP-TH 0208196;%%
}

%\LambertZR
\lref\LambertZR{
N.~Lambert, H.~Liu and J.~Maldacena,
``Closed strings from decaying D-branes,''
arXiv:hep-th/0303139.
%%CITATION = HEP-TH 0303139;%%
}

%\GaiottoRM
\lref\GaiottoRM{
D.~Gaiotto, N.~Itzhaki and L.~Rastelli,
``Closed strings as imaginary D-branes,''
arXiv:hep-th/0304192.
%%CITATION = HEP-TH 0304192;%%
}

%\SenMD
\lref\SenMD{
A.~Sen,
``Supersymmetric world-volume action for non-BPS D-branes,''
JHEP {\bf 9910}, 008 (1999)
[arXiv:hep-th/9909062].
%%CITATION = HEP-TH 9909062;%%
}

%\GarousiTR
\lref\GarousiTR{
M.~R.~Garousi,
``Tachyon couplings on non-BPS D-branes and Dirac-Born-Infeld action,''
Nucl.\ Phys.\ B {\bf 584}, 284 (2000)
[arXiv:hep-th/0003122].
%%CITATION = HEP-TH 0003122;%%
}

%\BergshoeffDQ
\lref\BergshoeffDQ{
E.~A.~Bergshoeff, M.~de Roo, T.~C.~de Wit, E.~Eyras and S.~Panda,
``T-duality and actions for non-BPS D-branes,''
JHEP {\bf 0005}, 009 (2000)
[arXiv:hep-th/0003221].
%%CITATION = HEP-TH 0003221;%%
}

%\KlusonIY
\lref\KlusonIY{
J.~Kluson,
``Proposal for non-BPS D-brane action,''
Phys.\ Rev.\ D {\bf 62}, 126003 (2000)
[arXiv:hep-th/0004106].
%%CITATION = HEP-TH 0004106;%%
}

%\SenTM
\lref\SenTM{
A.~Sen,
``Dirac-Born-Infeld action on the tachyon kink and vortex,''
Phys.\ Rev.\ D {\bf 68}, 066008 (2003)
[arXiv:hep-th/0303057].
%%CITATION = HEP-TH 0303057;%%
}

%\SenMV
\lref\SenMV{
A.~Sen,
``Remarks on tachyon driven cosmology,''
arXiv:hep-th/0312153.
%%CITATION = HEP-TH 0312153;%%
}

%\ElitzurPQ
\lref\ElitzurPQ{
S.~Elitzur, A.~Giveon, D.~Kutasov, E.~Rabinovici and G.~Sarkissian,
``D-branes in the background of NS fivebranes,''
JHEP {\bf 0008}, 046 (2000)
[arXiv:hep-th/0005052].
%%CITATION = HEP-TH 0005052;%%
}

%\PelcKB
\lref\PelcKB{
O.~Pelc,
``On the quantization constraints for a D3 brane in the geometry of NS5
%branes,''
JHEP {\bf 0008}, 030 (2000)
[arXiv:hep-th/0007100].
%%CITATION = HEP-TH 0007100;%%
}

%\RibaultSG
\lref\RibaultSG{
S.~Ribault,
``D3-branes in NS5-branes backgrounds,''
hep-th/0301092,
JHEP {\bf 0302}, 044 (2003)
.
%%CITATION = HEP-TH 0301092;%%
}

%\SeibergZK
\lref\SeibergZK{
N.~Seiberg,
``New theories in six dimensions and matrix description of M-theory on  T**5
%and T**5/Z(2),''
Phys.\ Lett.\ B {\bf 408}, 98 (1997)
[arXiv:hep-th/9705221].
%%CITATION = HEP-TH 9705221;%%
}

%\DeWolfeQX
\lref\DeWolfeQX{
O.~DeWolfe, S.~Kachru and H.~Verlinde,
``The giant inflaton,''
arXiv:hep-th/0403123.
%%CITATION = HEP-TH 0403123;%%
}

%\AlishahihaEH
\lref\AlishahihaEH{
M.~Alishahiha, E.~Silverstein and D.~Tong,
``DBI in the sky,''
arXiv:hep-th/0404084.
%%CITATION = HEP-TH 0404084;%%
}

%\BuchelQG
\lref\BuchelQG{
A.~Buchel and A.~Ghodsi,
``Braneworld inflation,''
arXiv:hep-th/0404151.
%%CITATION = HEP-TH 0404151;%%
}

%\GutperleAI
\lref\GutperleAI{
M.~Gutperle and A.~Strominger,
``Spacelike branes,''
JHEP {\bf 0204}, 018 (2002)
[arXiv:hep-th/0202210].
%%CITATION = HEP-TH 0202210;%%
}

%\GibbonsMD
\lref\GibbonsMD{
G.~W.~Gibbons,
``Cosmological evolution of the rolling tachyon,''
Phys.\ Lett.\ B {\bf 537}, 1 (2002)
[arXiv:hep-th/0204008].
%%CITATION = HEP-TH 0204008;%%
}

%\ShiuXP
\lref\ShiuXP{
G.~Shiu, S.~H.~H.~Tye and I.~Wasserman,
``Rolling tachyon in brane world cosmology from superstring field theory,''
Phys.\ Rev.\ D {\bf 67}, 083517 (2003)
[arXiv:hep-th/0207119].
%%CITATION = HEP-TH 0207119;%%
}

%\GiveonZM
\lref\GiveonZM{
A.~Giveon, D.~Kutasov and O.~Pelc,
``Holography for non-critical superstrings,''
JHEP {\bf 9910}, 035 (1999)
[arXiv:hep-th/9907178].
%%CITATION = HEP-TH 9907178;%%
}

%\AharonyUB
\lref\AharonyUB{
O.~Aharony, M.~Berkooz, D.~Kutasov and N.~Seiberg,
``Linear dilatons, NS5-branes and holography,''
JHEP {\bf 9810}, 004 (1998)
[arXiv:hep-th/9808149].
%%CITATION = HEP-TH 9808149;%%
}

%\GiveonPX
\lref\GiveonPX{
A.~Giveon and D.~Kutasov,
``Little string theory in a double scaling limit,''
JHEP {\bf 9910}, 034 (1999)
[arXiv:hep-th/9909110].
%%CITATION = HEP-TH 9909110;%%
}

%\BurgessQV
\lref\BurgessQV{
C.~P.~Burgess, P.~Martineau, F.~Quevedo and R.~Rabadan,
``Branonium,''
JHEP {\bf 0306}, 037 (2003)
[arXiv:hep-th/0303170].
%%CITATION = HEP-TH 0303170;%%
}

%\SilversteinHF
\lref\SilversteinHF{
E.~Silverstein and D.~Tong,
``Scalar speed limits and cosmology: Acceleration from D-cceleration,''
arXiv:hep-th/0310221.
%%CITATION = HEP-TH 0310221;%%
}

%\KehagiasVR
\lref\KehagiasVR{
A.~Kehagias and E.~Kiritsis,
``Mirage cosmology,''
JHEP {\bf 9911}, 022 (1999)
[arXiv:hep-th/9910174].
%%CITATION = HEP-TH 9910174;%%
}

%\BurgessTZ
\lref\BurgessTZ{
C.~P.~Burgess, F.~Quevedo, R.~Rabadan, G.~Tasinato and I.~Zavala,
``On bouncing brane worlds, S-branes and branonium cosmology,''
JCAP {\bf 0402}, 008 (2004)
[arXiv:hep-th/0310122].
%%CITATION = HEP-TH 0310122;%%
}

%\LuninGW
\lref\LuninGW{
O.~Lunin, S.~D.~Mathur, I.~Y.~Park and A.~Saxena,
``Tachyon condensation and 'bounce' in the D1-D5 system,''
Nucl.\ Phys.\ B {\bf 679}, 299 (2004)
[arXiv:hep-th/0304007].
%%CITATION = HEP-TH 0304007;%%
}

%\MaldacenaSS
\lref\MaldacenaSS{
J.~M.~Maldacena, G.~W.~Moore and N.~Seiberg,
``D-brane charges in five-brane backgrounds,''
JHEP {\bf 0110}, 005 (2001)
[arXiv:hep-th/0108152].
%%CITATION = HEP-TH 0108152;%%
}

%\GibbonsHF
\lref\GibbonsHF{
G.~W.~Gibbons, K.~Hori and P.~Yi,
``String fluid from unstable D-branes,''
Nucl.\ Phys.\ B {\bf 596}, 136 (2001)
[arXiv:hep-th/0009061].
%%CITATION = HEP-TH 0009061;%%
}

%\GibbonsTV
\lref\GibbonsTV{
G.~Gibbons, K.~Hashimoto and P.~Yi,
``Tachyon condensates, Carrollian contraction of Lorentz group, and fundamental
%strings,''
JHEP {\bf 0209}, 061 (2002)
[arXiv:hep-th/0209034].
%%CITATION = HEP-TH 0209034;%%
}

%\YeeEC
\lref\YeeEC{
H.~U.~Yee and P.~Yi,
``Open / closed duality, unstable D-branes, and coarse-grained closed
%strings,''
Nucl.\ Phys.\ B {\bf 686}, 31 (2004)
[arXiv:hep-th/0402027].
%%CITATION = HEP-TH 0402027;%%
}

\rightline{EFI-04-16}
\Title{
}
{\vbox{\centerline{D-Brane Dynamics Near NS5-Branes}}}
\bigskip
\centerline{David Kutasov}
\bigskip
\centerline{{\it Enrico Fermi Inst. and Dept. of Physics,
University of Chicago}}
\centerline{\it 5640 S. Ellis Ave., Chicago, IL 60637-1433, USA}
\bigskip\bigskip\bigskip
\noindent
We use the Dirac-Born-Infeld action to study the real time 
dynamics of a $Dp$-brane propagating in the vicinity of 
$NS5$-branes. This problem is closely related to tachyon 
condensation on an unstable D-brane, with the role of the 
tachyon played by the radial mode on the D-brane. As the 
D-brane approaches the fivebranes, its equation of state 
approaches that of a pressureless fluid. The pressure goes 
to zero at late times like $\exp(-\alpha t)$; $\alpha$ is 
a function of the number of fivebranes and of the angular
momentum of the D-brane. For unstable D-branes a similar
equation of state is taken to signal the decay of the
D-brane into closed string radiation. We propose that in
our case the D-brane decays into modes propagating in the
fivebrane throat, and show that this is consistent with
spacetime expectations. We also argue that for radial
motions of the D-brane deep inside the  throat,
the rolling process is described by an exactly
solvable worldsheet conformal field theory.

\vfill

\Date{5/04}

%%%%%%%%%%%%%%%%%%%%%%%%%%%%%%%%%%%%%%%%%%%%%%%%%%%%%%%%%%%%%%%%%%%%%%
%%%%%%%%%%%%%%%%%%%%%%%%%%%%%%%%%%%%%%%%%%%%%%%%%%%%%%%%%%%%%%%%%%%%%%
\newsec{Introduction}

In studies of D-branes one often finds that the lowest
lying open string state is a tachyon. This happens, for
example, for non-BPS D-branes and $D-\bar D$ systems in
the superstring. The presence of the tachyon signals an
instability of the D-brane system. Its condensation leads
to a more stable brane configuration, or in some cases
to the total annihilation of the brane. The latter is
possible since the branes in question do not carry any
conserved charges.

If one displaces the tachyon from the maximum of its potential
corresponding to the original unstable brane, it will roll
towards a minimum of the potential. This time-dependent process
can, in some cases, be understood exactly in classical open string
theory, by Wick rotating known Euclidean solutions to Minkowski
spacetime \SenNU\ (see also \refs{\GutperleAI,\LarsenWC}). 
The process of Wick rotation involves some physical choices 
associated with the choice of the initial state. Thus, the 
study of real time tachyon condensation provides useful 
insights on time-dependent solutions in string theory in general.

Studies of rolling tachyon solutions suggest that, at late times,
tachyon condensation leads to a peculiar ``tachyon matter'' state,
which has an equation of state of a pressureless fluid, and no
on-shell open string excitations. Apriori, one would expect the
unstable D-brane to decay into closed string modes, see \eg\
\refs{\OkudaYD,\LambertZR,\GaiottoRM}. It has been suggested
that the open string picture of the time evolution provides an
alternative but equivalent description of this decay, leading to
a new kind of open-closed string duality \refs{\SenXS,\SenIV}.
Related work on tachyon condensation appears in
\refs{\GibbonsHF,\GibbonsTV,\YeeEC}.

Another outcome of the recent work on real time tachyon condensation
is the observation that an effective action of the Dirac-Born-Infeld
(DBI) type for the tachyon
\refs{\SenMD\GarousiTR\BergshoeffDQ-\KlusonIY}
captures surprisingly well many aspects of rolling tachyon solutions of
the full open string theory \refs{\SenIN,\SenAN, \SenTM,\LambertZR},
and is thus very useful for studying these processes. The origin of the
agreement was partially clarified in \refs{\KutasovER,\NiarchosRW}.
There was also much work on possible cosmological applications of
tachyon condensation (see \eg\  \refs{\GibbonsMD,\ShiuXP,\SenMV} and
references therein).

In this paper we will study another kind of instability of (BPS)
D-branes, associated with the presence of $NS5$-branes in their
vicinity. As we will see, this example is closely related to the
tachyon condensation problem, and one can say a lot about the
nature of the time-dependent solutions in it. Thus, it adds to
our understanding of time-dependence in string theory, a subject
that suffers from a dearth of solvable examples.

We will consider a stack of $k$ parallel $NS5$-branes
in type II string theory, stretched in the directions
$(x^1,\cdots, x^5)$ and localized in $\vec x=(x^6,x^7,x^8,x^9)$.
The directions along the worldvolume of the fivebranes will
be denoted by $x^\mu$, $\mu=0,1,2,\cdots, 5$; those transverse
to the branes will be labeled by $x^m$, $m=6,7,8,9$.

Our interest will be in the dynamics of a BPS $Dp$-brane in
the presence of the fivebranes\foot{As usual, $p$ must be
even for type IIA string theory, and odd in the IIB case
\PolchinskiRR.}. The D-brane is ``parallel'' to the fivebranes, 
\ie\ it is extended in some or all of the fivebrane worldvolume 
directions $x^\mu$, and pointlike in the directions
transverse to the fivebranes $(x^6,x^7,x^8,x^9$). Without loss
of generality, we can take the worldvolume of the $Dp$-brane to
fill the directions $(x^0,x^1,\cdots, x^p)$. We will label the
worldvolume of the $Dp$-brane by $x^\mu$ as well, but one should
bear in mind that here the index $\mu$ only runs over the range
$\mu=0,1,2,\cdots,p$, with $p\le 5$.

Although the D-brane in question is BPS, in the presence of the
fivebranes it is unstable. Indeed, the configuration
of parallel $NS5$-branes and $Dp$-brane breaks supersymmetry
completely, since the fivebranes and D-brane preserve different
halves of the supersymmetry of type II string theory. The $Dp$-brane
carries RR charge, but this charge can leak from the D-brane to
the fivebranes. A quick way to see that is to use U-duality to map
the brane configuration described above to one that contains a
wrapped fundamental IIB string parallel to a stack of $k$
$D5$-branes. Since the string can end on the fivebranes, it can
decay into a collection of open strings living on the fivebranes.
In the process, its Neveu-Schwarz $B_{\mu\nu}$ charge is
transferred to the $D5$-branes. S-duality relates this to the statement
that the RR charge carried by a D-string can be transferred to parallel
$NS5$-branes. T-duality along the fivebrane worldvolume
generalizes this statement to all $p$.

If we place a BPS $Dp$-brane at a finite distance from a stack
of $NS5$-branes, it will experience an attractive force, and will
start moving towards the fivebranes. There are a number of natural
questions one can ask about this dynamical process. One is what
happens as the $Dp$-brane approaches the $NS5$-branes. As we
discuss in section 4, we would expect the D-brane to shed most of
its energy and form a bound state with the fivebranes. Can this
process be studied in a controlled way?

Another interesting question is whether there are stable trajectories
where the D-brane is in orbit around the collection of fivebranes.
One might expect the answer to be negative, since
at least when the D-brane is far from the fivebranes in the transverse
space labeled by $\vec x$, it should experience a standard gravitational
potential $V\sim -1/|\vec x|^2$ (since the transverse space is four
dimensional) and this potential does not allow for stable orbits. But,
this leaves the question whether there are such orbits in the full
gravitational potential of the fivebranes, which deviates significantly
from $-1/|\vec x|^2$ at short distances.

We will see that these questions can be usefully addressed using weakly
coupled string theory techniques. This might sound surprising since it is
well known that coincident $NS5$-branes develop a throat along which the
string coupling grows without bound as one approaches the fivebranes.
A D-brane falling towards the fivebranes will eventually explore the
strong coupling region; hence, one expects string perturbation theory
to break down at late times. We will argue that in fact there is a range 
of energies of the D-brane, for which a significant part of the dynamical 
process occurs in the region where the string coupling is weak, and 
perturbative string theory is reliable. Furthermore, we will see that 
in some cases, the motion of the D-brane in the throat of the fivebranes 
can be described by an exactly solvable worldsheet CFT, which is obtained 
by Wick rotation of the supersymmetric generalization of the Euclidean 
D-brane constructed in \LukyanovNJ.

\newsec{The effective action for a D-brane 
in the $NS5$-brane background}

At weak string coupling, $NS5$-branes are much heavier 
than D-branes -- their tension goes like $1/g_s^2$, 
while that of D-branes goes like $1/g_s$. To study the 
dynamics of a D-brane in the vicinity of fivebranes in 
this regime, we can take the fivebranes to be static and 
study the motion of the D-brane in their gravitational 
potential. For related earlier work, see \eg\ 
\refs{\ElitzurPQ,\PelcKB,\RibaultSG}.

The background fields around $k$ $NS5$-branes are given 
by the CHS solution \refs{\CallanAT,\PolchinskiRR}. The 
metric, dilaton and NS $B$-field are
\eqn\chs{\eqalign{
&ds^2=dx_\mu dx^\mu+H(x^n) dx^mdx^m\cr
&e^{2(\Phi-\Phi_0)}=H(x^n)\cr
&H_{mnp}=-\epsilon_{mnp}^q\partial_q\Phi~.\cr
}}
Here $H(x^n)$ is the harmonic function describing $k$
fivebranes, and $H_{mnp}$ is the field strength of the
NS $B$-field. For fivebranes at generic positions
$\vec x_1,\cdots, \vec x_k$, one has
\eqn\hgen{H=1+l_s^2\sum_{j=1}^k
{1\over |\vec x-\vec x_j|^2}~,}
where $l_s=\sqrt{\alpha'}$ is the string length.
We will mainly discuss the case of coincident fivebranes,
where an $SO(4)$ symmetry group of rotations around the
fivebranes is preserved, and the harmonic function \hgen\
reduces to
\eqn\hcoin{H=1+{kl_s^2\over r^2}~.}
$r=|\vec x|$ is the radial coordinate away from the fivebranes
in the transverse $\IR^4$ labeled by $(x^6,\cdots, x^9)$.

To construct D-branes in the fivebrane background, one has to
study the sigma model with the target space fields \chs\ on a
worldsheet with boundary, and determine possible boundary
conditions which preserve conformal invariance.

In this paper we will study this problem using an effective action on
the worldvolume of the D-brane, the DBI action. This will lead to an
approximate treatment that, as mentioned above, is known to be quite
reliable. We will comment on the full conformal field theory problem
below, leaving a more detailed discussion to another publication.

As mentioned above, we will study a $Dp$-brane stretched in the
directions $(x^1,\cdots,x^p)$. We can label the worldvolume
of the D-brane by $\xi^\mu$, $\mu=0,1,2,\cdots, p$, and use
reparametrization invariance on the worldvolume of the D-brane
to set $\xi^\mu=x^\mu$. The position of the D-brane in the transverse
directions $(x^6,\cdots, x^9)$ gives rise to scalar fields on the
worldvolume of the D-brane, $(X^6(\xi_\mu),\cdots, X^9(\xi_\mu))$.
The dynamics of these fields is governed by the DBI action
\refs{\PolchinskiRQ,\TseytlinDJ}
\eqn\dbi{\SS_p=-\tau_p\int d^{p+1}\xi e^{-(\Phi-\Phi_0)}
\sqrt{-\det(G_{\mu\nu}+B_{\mu\nu})}~.}
$\tau_p$ is the tension of the $Dp$-brane, and the determinant
runs over the worldvolume directions, $\mu=0,1,\cdots, p$.
$G_{\mu\nu}$ and $B_{\mu\nu}$ are the induced metric
and $B$-field on the D-brane:
\eqn\gb{\eqalign{
G_{\mu\nu}={\partial X^A\over\partial \xi^\mu}
{\partial X^B\over\partial \xi^\nu}G_{AB}(X)~,\cr
B_{\mu\nu}={\partial X^A\over\partial \xi^\mu}
{\partial X^B\over\partial \xi^\nu}B_{AB}(X)~.\cr
}}
The indices $A,B=0,1,2,\cdots, 9$ run over the whole
ten dimensional spacetime. $G_{AB}$ and $B_{AB}$ are the
metric and $B$-field in ten dimensions. In our case they
are given by \chs. There is also a gauge field on the
D-brane that one can turn on, but we will not do that.

The action \dbi\ is expected to be reliable for arbitrary 
values of the gradients $\partial_\mu X^m$, as long as they 
are slowly varying in spacetime, in a suitable sense. In 
addition, the string coupling at the location of the brane 
should be small, \ie\ one must have $\exp(\Phi)<<1$ in \chs. 
We will comment on the reliability of this action further below.

An interesting special case of the action \dbi\ is obtained
by placing all the $NS5$-branes at $\vec x=0$, and restricting
to purely radial fluctuations of the $Dp$-brane in the transverse
$\IR^4$ labeled by $\vec x$. For such fluctuations, the only field
on the brane that is excited is $R(\xi_\mu)=\sqrt{X^mX^m(\xi_\mu)}$;
the angular variables remain fixed at their initial values. This
restriction to radial motion is consistent since, for coincident
$NS5$-branes, the background \chs\ is $SO(4)$ invariant, and
the D-brane experiences a central force pulling it towards
the origin.

The action \dbi\ takes a rather simple form in this case.
Since the $B$-field \chs\ is in the angular directions,
and the angular degrees of freedom are not excited,
the induced $B$-field in \dbi\ vanishes. The induced
metric \gb\ takes the form
\eqn\indmetr{G_{\mu\nu}=\eta_{\mu\nu}+
H(R)\partial_\mu R\partial_\nu R~.}
The DBI action \dbi\ is thus given by
\eqn\dbiradial{\SS_p=-\tau_p\int d^{p+1}x
{1\over\sqrt H}\sqrt{-\det G_{\mu\nu}}
=-\tau_p\int d^{p+1}x{1\over\sqrt H}
\sqrt{1+H(R)\partial_\mu R\partial^\mu R}~.}
This action is very reminiscent of the DBI action for
the tachyon in open string models
\refs{\SenMD\GarousiTR\BergshoeffDQ\KlusonIY\SenIN\SenAN\SenTM-\KutasovER} ,
\eqn\dbitachyon{\SS_{\rm tach}=
-\int d^{p+1}x V(T)\sqrt{-\det G^{\rm (tach)}_{\mu\nu}}
=-\int d^{p+1}x V(T)\sqrt{1+\partial_\mu T\partial^\mu T}}
where
\eqn\indmetrtach{G^{\rm (tach)}_{\mu\nu}=
\eta_{\mu\nu}+\partial_\mu T\partial_\nu T~,}
and $V(T)$ is the tachyon potential. Comparing \indmetr\
and \indmetrtach\ we see that we can map one to the other as
follows. Define a ``tachyon'' field $T$ via the relation
\eqn\tachdef{{dT\over dR}=\sqrt{H(R)}=\sqrt{1+{kl_s^2\over R^2}}~.}
In terms of this field, the induced metric \indmetr\ becomes the 
same as \indmetrtach,
and the ``tachyon potential'' $V(T)$ in \dbitachyon\ is given by
\eqn\effpot{V(T)={\tau_p\over\sqrt {H(R(T))}}~.}
The solution of eq. \tachdef\ is 
(up to an unimportant additive constant)
\eqn\soltach{T(R)=\sqrt{kl_s^2+R^2}+\half\sqrt{k}l_s
\ln{\sqrt{kl_s^2+R^2}-\sqrt{k}l_s\over\sqrt{kl_s^2+R^2}+\sqrt{k}l_s}~.}
{}From \tachdef\ we see that $T$ is a monotonically increasing function
of $R$. As $R\to 0$, $T(R)\to-\infty$: 
\eqn\trzero{T(R\to0)\simeq \sqrt{k}l_s \ln {R\over\sqrt kl_s}~.}
As $R\to\infty$, $T\to\infty$:
\eqn\trinfty{T(R\to\infty)\simeq R~.}
We can use this to analyze the behavior of the effective potential for $T$,
\effpot, in the different limits. One finds
\eqn\vefftt{{1\over\tau_p}V(T)\simeq
\cases{\exp{T\over\sqrt{k}l_s}&$T\to-\infty$\cr
1-{kl_s^2\over 2T^2}&$T\to\infty $~.\cr
}}
As $T\to\infty$, we recognize the long range gravitational attraction
between the D-brane and the fivebranes. Indeed, the interaction
potential \vefftt\ goes like $-1/T^2\simeq-1/R^2$. In the opposite
regime, $T\to-\infty$, corresponding to $R\to 0$, we find that the
potential $V(T)$ \vefftt\ goes exponentially to zero. This is precisely
the behavior exhibited at late times by the tachyon potential relevant
for rolling tachyon solutions. This behavior of the potential leads to the
absence of plane wave solutions around the minimum of the potential at
$T\to-\infty$, and to the exponential decrease of the pressure at late times
\SenAN. In our case, the ``tachyon field'' $T$ which appears in the DBI
Lagrangian \dbitachyon\ acquires a geometric meaning -- it is related
via \soltach\ to the distance between the D-brane and the fivebranes.
Note that for $k=2$ (two fivebranes) the slope of the exponential in
\vefftt\ is the same as that for non-BPS D-branes in type II string theory.

The radial DBI action \dbiradial\ interpolates smoothly between 
standard gravitational attraction of the D-brane to the fivebranes 
at large distances and a ``radion matter'' phase when the $Dp$-brane 
is close to the fivebranes. The transition between the two 
behaviors occurs at $R\sim\sqrt{k}l_s$. Note that even when 
the D-brane is very close to the fivebranes, there is no 
perturbative string tachyon here -- a fundamental 
string cannot stretch between the D-brane and the 
$NS5$-branes, since it cannot end on the latter.

In fact, one would naively expect that the regime where the
D-brane is approaching the $NS5$-branes cannot be described
by perturbative string theory at all, since the string coupling
is large there. In the next section we will study some 
solutions of the equations of motion of the DBI action, and
will see that in many cases, the open string description of 
the collapse of the D-brane onto the fivebranes is reliable 
up to rather late times, long after the D-brane collapsed 
into an effectively pressureless fluid.

It is natural to ask what happens when instead of a stack
of $NS5$-branes, the $Dp$-brane approaches a stack of other
D-branes. A case that is closely related to our discussion
is that of a $D3$-brane stretched in $(x^1,x^2,x^3)$ 
approaching a stack of $k$ $D5$-branes stretched in 
$(x^1,\cdots,x^5)$. This system is S-dual to the one 
discussed here for $p=3$. S-duality implies that the DBI
action on the $D3$-brane has the same form as \dbiradial;
the only difference is that since under S-duality
$l_s^2\to l_s^2g_s$, the harmonic function of the 
fivebranes, \hcoin, is modified to
\eqn\hcoindd{H_D=1+{kg_sl_s^2\over r^2}~.}
In this case, there are two rather different regimes.
For $kg_s<<1$ the description in terms of the metric
\chs, \hcoindd\ is inappropriate, and one should study
the system using open strings ending on the $D5$-branes.
In particular, when the $D3$-brane comes to within a distance
$l_s$ from the $D5$-branes, a tachyon appears in the spectrum
and the dynamics of the system is governed by its condensation
from that point on. 

For $kg_s>>1$, the discussion is rather similar to the $NS5$-brane
case. The description in terms of the closed string background
\chs, \hcoindd\ is good and, as above, when the distance between
the $D3$-brane and the $D5$-branes $R$ is much smaller than
$\sqrt{kg_s}l_s$ (but still, perhaps, much larger than $l_s$), 
the radial mode $R(x^\mu)$ is described by the ``tachyon matter''
Lagrangian \dbitachyon\ with an exponentially decaying potential
given by \vefftt\ (with $l_s\to l_s\sqrt{g_s}$). The main advantage
of the $NS5$-brane system is that since the closed string background
\chs\ does not involve RR fields, it is possible to study exact D-brane
solutions, something that is difficult in the $D5$-brane background.

If one replaces the $k$ $D5$-branes by $Dp$-branes with other values
of $p$, the relevant harmonic function no longer goes like $1/r^2$ for
small $r$. Thus, the relation \trzero\ between the radial mode and the
analog of the field $T$ is no longer logarithmic, but rather
powerlike. Also, in some cases one has to include the contribution
of the RR field of the $k$ $Dp$-branes to the Lagrangian of the
probe D-brane. We will not discuss this in detail here; see \eg\
\refs{\BurgessQV,\LuninGW,\SilversteinHF} for some recent papers
and further references.

\newsec{Solutions of the equations of motion}

In this section we would like to study some solutions of
the equations of motion of the DBI action \dbi\ for the
case of coincident  $NS5$-branes, \hcoin. We
will mostly consider here the homogeneous case, where
the D-brane collapses towards the fivebranes uniformly,
but before specializing to this case we would like to
comment briefly on inhomogeneous solutions.

In the case of radial motions \dbiradial, we can use the
map to the tachyon problem \dbitachyon, \effpot, \soltach\
to draw on some recent studies of classical solutions for
Lagrangians of the general form \dbitachyon.
For late times the solutions approach the surface \FelderSV
\eqn\latesurf{\partial_\mu T\partial^\mu T=-1}
or in terms of the original variable $R(x^\mu)$ \dbiradial, 
\tachdef,
\eqn\maxvel{H(R)\partial_\mu R\partial^\mu R=-1~.}
The solutions of \latesurf\ can be studied analytically
and exhibit a rich structure \refs{\FelderSV,\FelderXU}.
Any inhomogeneities in the initial data grow with time, 
and the solutions develop caustics at a finite time. As 
one approaches these caustics, the analysis based on the
effective action \dbiradial\ breaks down and has to 
be replaced by a full CFT analysis.

In the rest of this section, we will study homogeneous
solutions of the equations of motion of the DBI action
\dbi, for which the fields representing the transverse
position of the D-brane, $X^m$, $m=6,7,8,9$ depend only
on time,  $X^m=X^m(t)$, and the caustics mentioned above
do not form. In this case, the action \dbi\ simplifies. 
The induced $B$-field \gb\ vanishes, and the induced 
metric takes the form
\eqn\indghom{G_{\mu\nu}=\eta_{\mu\nu}+
\delta_\mu^0\delta_\nu^0 \dot X^m\dot X^m H(X^n)~.}
Substituting the  metric \indghom\ and dilaton \chs\ into the
action \dbi\ we find the following action for $X^m(t)$:
\eqn\seff{\SS_p=-\tau_p V\int dt{1\over\sqrt{H(X^n)}}
\sqrt{1-\dot X^m\dot X^m H(X^n)}=
-\tau_p V\int dt\sqrt{H(X^n)^{-1}-\dot X^m\dot X^m}~,}
where $V$ is the volume of the $p$ dimensional space
the D-brane is stretched in. In the case with
no fivebranes, $H=1$, and \seff\ reduces to the 
familiar action for a D-brane in flat spacetime.

The equations of motion of the Lagrangian \seff\ take the
form:
\eqn\eom{{d\over dt}\left({\dot X^n\over
\sqrt{H^{-1}-\dot X^m\dot X^m}}\right)=
{\partial_n H\over 2H^2\sqrt{H^{-1}-\dot X^m\dot X^m}}~.}
There are also some conserved charges in this system.
Time translation invariance  implies that the energy
\eqn\eee{E=P_n \dot X^n-\LL}
is conserved. The momentum $P_n$ is obtained by varying the
Lagrangian $\LL$,
\eqn\ppii{P_n={\delta \LL\over \delta \dot X^n}=
{\tau_p V\dot X_n\over \sqrt{H^{-1}-\dot X^m\dot X^m}}~.}
Substituting \ppii\ into \eee\ we find that the energy is given by
\eqn\eeee{E={\tau_p V\over H\sqrt{H^{-1}-\dot X^m\dot X^m}}~.}
One can check directly that $E$ \eeee\ is conserved by multiplying
the equations of motion \eom\ by $\dot X^n$ and summing over $n$.

So far, we discussed the case of general $H$, or fivebranes at
generic locations, $\vec x_i$, in the transverse $\IR^4$.
Now, we would like to specialize to the case of coincident
fivebranes. Thus, we will take $H$ to only depend on
$R=\sqrt{X^mX^m}$, $H=H(R)$, \hcoin. The equations
of motion \eom\ take in this case the form
\eqn\eomcoin{{d\over dt}\left({\dot X^n\over
\sqrt{H^{-1}-\dot X^m\dot X^m}}\right)=
{X^n H'\over 2RH^2\sqrt{H^{-1}-\dot X^m\dot X^m}}~.}
To solve these equations we need to specify  initial
data $\vec X(t=0)$ and $\dot{\vec X}(t=0)$. These two vectors
define a plane in $\IR^4$. By an $SO(4)$ rotation
-- a symmetry of the problem -- we can rotate this plane into,
say, the $(x^6,x^7)$ plane. Then the motion will remain in the 
$(x^6,x^7)$ plane for all time. Thus, without loss of generality, 
we can study trajectories in this plane. 

In addition to the energy, the angular momentum of
the D-brane is conserved as well. It is given by
\eqn\llll{L=X^6P^7-X^7P^6~.}
Using the expression for the momentum, \ppii, we find that
\eqn\lnew{L=\tau_p V
{X^6\dot X^7-X^7\dot X^6\over\sqrt{H^{-1}-\dot X^m\dot X^m}}~.}
One can check directly that $L$ \lnew\ is conserved
by using the equations of motion \eomcoin.

Another quantity of interest is the stress tensor $T_{\mu\nu}$
associated with the moving D-brane. $T_{00}$ is the
energy density, so it is given by our expression for $E$, \eeee,
with the factor of the volume stripped off.
The space-space components of $T_{\mu\nu}$ can be calculated
by the usual Noether procedure. One finds $(i,j=1,2,\cdots,p)$
\eqn\pressure{\eqalign{
T_{00}=&{\tau_p\over H\sqrt{H^{-1}-\dot X^m\dot X^m}}~,\cr
T_{ij}=&-\tau_p\delta_{ij} \sqrt{H^{-1}-\dot X^m\dot X^m}~,\cr
T_{0i}=&0~.\cr}}
The conservation equations \eeee, \lnew\ can be integrated to find
the possible trajectories of the D-brane in the $(x^6,x^7)$ plane. As
usual in central potential problems, it is convenient to pass to polar
coordinates,
\eqn\polar{\eqalign{
X^6=&R\cos\theta~,\cr
X^7=&R\sin\theta~.\cr
}}
In these coordinates we have the following expressions for the energy
density and for the angular momentum density (which we will denote
by the same letters as the energy \eeee\ and angular momentum \lnew):
\eqn\polarel{\eqalign{
E=&{\tau_p\over H\sqrt{H^{-1}-\dot R^2-R^2\dot\theta^2}}~,\cr
L=&\tau_p R^2\dot\theta\over\sqrt{H^{-1}-\dot R^2-R^2\dot\theta^2}~.\cr
}}
In order to solve the equations of motion for given energy and
angular momentum densities $E$ and $L$, we would like to
solve the second equation in \polarel\ for $\dot\theta$ and
substitute the solution in the first one. The solution for
$\dot\theta$ is
\eqn\thetadot{\dot\theta^2={L^2(H^{-1}-\dot R^2)\over
\tau_p^2 R^4+L^2 R^2}~.}
Substituting it into \polarel\ and solving for $\dot R$, we find
\eqn\rdot{\dot R^2=
{1\over H}-{1\over E^2H^2}\left(\tau_p^2+{L^2\over R^2}\right)~.}
We would next like to study the solutions of the equations of motion
\thetadot, \rdot.

Consider first the case of vanishing angular momentum, $L=0$.
Eq. \thetadot\ implies that $\theta$ is constant, while the radial
equation \rdot\ takes the form
\eqn\rdotlzero{\dot R^2={1\over H}-{\tau_p^2\over E^2H^2}~.}
The solution is restricted to the region in which the right hand
side of \rdotlzero\ is non-negative. Substituting the form of $H$,
\hcoin, we find the constraint on $R$ (for fixed energy density $E$)
\eqn\restr{{kl_s^2\over R^2}\ge{\tau_p^2\over E^2}-1~.}
This equation has an obvious interpretaion. If the energy density
$E$ is larger than the tension of a free BPS $Dp$-brane, $\tau_p$,
the constraint \restr\ is empty and the $Dp$-brane can escape to infinity.
For $E<\tau_p$, the D-brane does not have enough energy to escape
the gravitational pull of the $NS5$-branes, and it cannot exceed some
maximal distance from them.

It is interesting to calculate the energy momentum tensor of
the D-brane for such a radial motion. The energy density
$T_{00}$ is constant and equal to $E$ throughout the time 
evolution, whereas for $T_{ij}$ (second line of \pressure) 
one finds
\eqn\pprree{T_{ij}=-\tau_p\delta_{ij}\sqrt{H^{-1}-\dot R^2}
=-\delta_{ij}{\tau_p^2\over E H(R)}~.}
We see that the pressure goes smoothly to zero as $R\to 0$, 
since $H(R)\sim 1/R^2$ there \hcoin.

If the entire trajectory $R(t)$ \rdotlzero\ remains
in the small $R$ region, such that $R<<\sqrt{k} l_s$
for all times (this happens if $\tau_p/E>>1$, see \restr),
we can solve for the trajectory $R(t)$ exactly. Indeed, in that
case one has $H=kl_s^2/R^2$; substituting this into \rdotlzero\
we find the equation of motion
\eqn\smallreom{\dot R^2={R^2\over kl_s^2}-{\tau_p^2 R^4\over
E^2 k^2l_s^4}~,}
with the solution
\eqn\solradial{{1\over R}=
{\tau_p\over E\sqrt{k} l_s}\cosh{t\over \sqrt{k} l_s}~,}
where we chose $t=0$ to be the time at which the D-brane 
reaches its maximal distance from the fivebranes. 
Substituting \solradial\ into \chs\ we find that the
dilaton behaves as a function of time as follows:
\eqn\philocal{e^\Phi={g_s\tau_p\over E}
\cosh{t\over\sqrt{k}l_s}~.}
Quantum effects are small when $\exp\Phi<<1$. 
Thus, we see that if the energy density lies in the range 
\eqn\enconst{g_s\tau_p<<E<<\tau_p~,}
there is a long period of time during which the solution
\solradial\ is reliable. 

The physics of $NS5$-branes is usually studied by taking 
the Little String Theory (LST) limit $g_s\to 0$ in the 
geometry \chs, \AharonyUB. Recalling that the $Dp$-brane tension 
has the form $\tau_p=T_p/g_s$, where $T_p\sim m_s^{p+1}$ is 
independent of $g_s$, we see that the constraints \enconst\ 
simplify in this regime, to $E>>m_s^{p+1}$. 

This is familiar from other studies of LST (such as \GiveonPX). 
The perturbative expansion of LST is an expansion in the inverse
of the tension of a D-brane localized in the throat; in this
case, it is a $1/E$ expansion \philocal. Non-perturbative effects
become important when $\cosh(t/\sqrt{k}l_s)\sim E/m_s^{p+1}$. 
For any given time, they can be made arbitrarily small 
by increasing $E$. 

In the throat of the fivebranes, it is natural \CallanAT\ to
perform the coordinate transformation
\eqn\coor{R=\sqrt{k}l_s e^{\phi\over\sqrt{k} l_s}~,}
since in terms of $\phi$ the metric \chs\ is 
$ds^2=d\phi^2+\cdots$, and the
dilaton is linear, $\Phi\simeq -\phi/\sqrt{k}l_s$.
Substituting \coor\ into \solradial, the solution 
takes the form
\eqn\solphi{e^{-{\phi\over\sqrt{k} l_s}}=
{\tau_p\over E}\cosh{t\over \sqrt{k} l_s}~.}
At very early and late times, $|t|\to\infty$,
one has 
\eqn\asympsol{\phi(t)\simeq -|t|~.}
Thus the D-brane approaches the speed of light
as $t\to\pm\infty$.

Several comments are appropriate at this point:
\item{(1)} While the full solution \solradial\ is
only valid in the regime \enconst, its asymptotic
late (or early) time behavior \asympsol\ is valid 
for all values of $E/\tau_p$, for solutions 
describing an infalling (or outgoing) D-brane, as
long as the string coupling \philocal\ remains small.
\item{(2)} In terms of the time $t$ seen by an observer
living on the $NS5$-branes, it takes an infinite amount 
of time for the D-brane to reach $R=0$. In the induced 
metric on the D-brane, \indghom, one has
\eqn\gooind{G_{00}=-1+\dot R^2H(R)=-
{\tau_p^2\over E^2 H(R)}=
-{1\over\cosh^2{t\over\sqrt{k}l_s}}~.}
Thus, the D-brane reaches the $NS5$-branes in
a finite proper time $\tau$, which is related to
$t$ via $-d\tau^2=G_{00}(t)dt^2$, or
\eqn\proptau{\tan{\tau\over 2\sqrt{k}l_s}=
e^{t\over\sqrt{k}l_s}~.}
In terms of $\tau$, the solution \solradial\ can
be written as
\eqn\soltau{R={E\over\tau_p}\sqrt{k}l_s\sin{\tau\over\sqrt{k}l_s}~.}
It describes an oscillatory motion, in which the 
$Dp$-brane goes from one side of the fivebranes 
to the other and back, going right through them
twice in each cycle. An observer on the D-brane
sees a flat geometry, $ds^2=-d\tau^2+dx^idx^i$,
and stress tensor 
\eqn\tbrane{\eqalign{
&T_{\mu\nu}^{({\rm brane})}=-\Lambda \eta_{\mu\nu}~,\cr
&\Lambda={\tau_p^2\over EH(R)}=E\sin^2{\tau\over\sqrt{k}l_s}~.
}}
Of course, the full solution \soltau\ may not be
reliable, since every time $R$ passes through
zero, the D-brane passes a region of strong coupling,
and the analysis based on weakly coupled string theory
breaks down. 
\item{(3)} The coordinate transformation \coor\
is also natural from the point of view of the analogy
to tachyon condensation discussed in section 2. Comparing
\trzero\ and \coor\ we see that deep inside the fivebrane
throat we can identify the tachyon $T$ with $\phi$, the
natural radial coordinate in the throat.
\item{(4)} Substituting the form of the solution \solradial\
into the expression for $T_{ij}$, \pprree, we find that the
pressure behaves at late times as
\eqn\preslate{P\sim \exp\left(-{2t\over\sqrt{k}l_s}\right)~.}
Both the exponential behavior, and the
coefficient in the exponent agree with the analysis of
\SenAN, given the fact that the effective potential $V(T)$
\vefftt\ behaves near its minimum at $T\to-\infty$ as
$V\sim \exp(T/\sqrt{k}l_s)$.
\item{(5)}If we Wick rotate the solution \solphi\ from Minkowski
to Euclidean space, replacing $t\to iY$, we get a static brane 
whose shape is given by
\eqn\euclbrane{e^{-{\phi\over\sqrt{k} l_s}}=
\alpha\cos{Y\over \sqrt{k} l_s}~,}
where $\alpha$ is an arbitrary constant. This brane starts out
at large positive $\phi$ (which can be thought of as the region
far from the fivebranes in the near-horizon geometry) and extends 
to some finite $\phi$ (determined by $\alpha$ \euclbrane) and back,
while curving in the $Y$ direction. The fact that this brane solves
the equations of motion is due to a competition between the linear dilaton,
which pulls the D-brane to $\phi\to-\infty$, and the energy cost associated
with the bending of the D-brane in the $Y$ direction. Note that the
string coupling everywhere on the D-brane \euclbrane\ can be made
arbitrarily small by tuning $\alpha$. Interestingly, in the
recent paper \LukyanovNJ, an exact boundary state for the Euclidean brane
\euclbrane\ was constructed, in the bosonic string. Assuming that the results
of \LukyanovNJ\ can be generalized to the superstring, we can use them to
describe the Minkowski D-brane \solphi\ by continuing the boundary state,
as in \SenNU. This would also provide further evidence for the fact
that the D-brane \solphi\ can be studied using perturbative string 
techniques, as the Euclidean D-brane of \LukyanovNJ\ is manifestly
non-singular.

\noindent
So far we discussed purely radial trajectories with vanishing angular
momentum density \polarel. A natural question is whether anything
qualitatively new  occurs for non-zero $L$. One way of thinking about
this is the following. The radial equation of motion \rdot\ 
can be thought of as describing a particle with mass $m=2$, 
moving in one dimension (labeled by $R$) in the potential
\eqn\effecpoten{V_{\rm eff}(R)=
{1\over E^2H^2}\left(\tau_p^2+{L^2\over R^2}\right)
-{1\over H},}
with zero energy.
The potential $V_{\rm eff}$ has the following features.
The leading behavior for small $R$ is
\eqn\veffsmall{V_{\rm eff}(R)\simeq
{R^2\over kl_s^2}\left({L^2\over E^2 kl_s^2}-1\right)~.}
For large $R$, we have instead
\eqn\vefflarge{V_{\rm eff}(R)\simeq {\tau_p^2\over E^2}-1~.}
If the energy density of the D-brane is smaller than that of a free
$Dp$-brane, $E<\tau_p$, such that it cannot escape to infinity,
$V_{\rm eff}$ approaches a positive constant \vefflarge\ as $R\to\infty$.
It is then easy to show that in order to have trajectories at non-zero
$R$, the angular momentum must satisfy the bound\foot{The fact that
there is an upper bound on $L$ is familiar from the study of
central potentials that go like $-1/r^2$ in Newtonian mechanics;
the precise form of the bound on $L$ relies on the full
non-linear structure of the DBI Lagrangian.}
\eqn\boundangu{L<\sqrt{k}l_sE~.}
If \boundangu\ is not satisfied, the only solution is $R=0$.
If \boundangu\ is satisfied, the trajectory of the D-brane is
qualitatively similar to that in the $L=0$ case. In particular, 
it approaches the fivebranes (\ie\ $R=0$) both at very 
early and very late time, and does not have stable orbits
at finite $R$. 

For $\tau_p>>E$, the whole trajectory lies again in the
region $R<<\sqrt{k}l_s$, and one can approximate the 
harmonic function \hcoin\ by $H=kl_s^2/R^2$. The equation
for $\dot R$, \rdot, takes a form analogous to \smallreom,
\eqn\smalllt{\dot R^2={R^2\over kl_s^2}
\left(1-{L^2\over kl_s^2 E^2}\right)
-{\tau_p^2 R^4\over E^2 k^2l_s^4}~,}
with the solution
\eqn\solraddll{{1\over R}=
{\tau_p\over \sqrt{k}l_sE\sqrt{1-{L^2\over kl_s^2 E^2}}}
\cosh\left({t\over \sqrt{k} l_s}
\sqrt{1-{L^2\over kl_s^2 E^2}}\right)~.}
We see that non-zero angular momentum slows down the
exponential decrease of $R$ as $t\to\infty$. From eq. 
\polarel\ we see that 
\eqn\thgen{\dot\theta={L\over E R^2H(R)}~.}
In the throat approximation one has 
$R^2H(R)=kl_s^2$. Thus, the solution is
\eqn\ththroat{\theta={L\over Ekl_s^2}t~.}
The solution \solraddll, \ththroat\ describes the
D-brane spiralling towards the origin, circling 
around it an infinite number of times in the 
process.

The pressure corresponding to the solution 
\solraddll\ can be read off \pressure:
\eqn\pressll{P=-\tau_p\sqrt{H^{-1}-\dot R^2-R^2\dot\theta^2}
=-{\tau_p^2\over EH(R)}=-{\tau_p^2R^2\over kl_s^2E},}
as in \pprree. Thus, it again goes exponentially
to zero, as in \preslate, with a tunable coefficient 
(which depends on $L$).  

For $E>\tau_p$, the situation is different. When the angular
momentum satisfies the bound \boundangu, there are two kinds
of trajectories. One corresponds to a D-brane that starts at 
$R=0$ at $t=-\infty$, and escapes to $R=\infty$ at late times;
the other describes the time reverse process.

If the angular momentum does not satisfy the bound
\boundangu, the D-brane starts at $R=\infty$ at early
times and ends there at late times, never getting
closer than the radius of closest approach, $R_{\rm min}$,
at which the effective potential $V_{\rm eff}$ \effecpoten\
vanishes,
\eqn\rrmin{R_{\rm min}^2={L^2-kl_s^2E^2\over
E^2-\tau_p^2}~.}
In this case, a natural physical question concerns the 
scattering of a D-brane off the fivebranes.
A D-brane with initial velocity $v$ and impact parameter 
$s$ has angular momentum and energy density
\eqn\lescat{\eqalign{
L=&{\tau_p sv\over\sqrt{1-v^2}}~,\cr
E=&{\tau_p\over\sqrt{1-v^2}}~.\cr
}}
Eliminating $v$, we find the relation
\eqn\angimp{L=s\sqrt{E^2-\tau_p^2}~.}
The D-brane interacts with the fivebranes
and goes back out to infinity after being 
deflected by the angle 
\eqn\defld{\delta\theta=\pi\left(
{1\over\sqrt{1-{kl_s^2E^2\over L^2}}}-1\right)~.}
As $L\to\infty$, the deflection angle \defld\ goes to 
zero, while as we approach the critical value \boundangu,
it diverges. This means that as $L\to \sqrt{k}l_sE$,
the D-brane circles around the fivebranes more and 
more times before eventually escaping to infinity. 
As mentioned above, for $L<\sqrt{k}l_sE$ the D-brane 
never makes it back to infinity. 

In terms of the impact parameter $s$, the bound 
\boundangu\ is equivalent via \angimp\ to 
\eqn\condss{s<{\sqrt{k}l_s\over \sqrt{1-{\tau_p^2\over E^2}}}~.}
If this condition is satisfied, the incoming D-brane is absorbed
by the fivebranes; otherwise it escapes back to infinity. 
We see that no matter how high the energy of the D-brane is,
if its impact parameter satisfies
\eqn\boundimpact{s<\sqrt{k}l_s~,}
it will be absorbed by the fivebranes. 
Thus, we see that when we probe a collection of $k$ fivebranes 
by scattering D-branes off it, this system behaves like an 
absorbing ball of radius $\sqrt{k}l_s$. One can show that the
same is true if we probe the fivebrane system by particles 
that move on geodesics in the metric \chs.

\newsec{Discussion}

As we saw in the previous sections, the real time dynamics
of a $Dp$-brane in the background of parallel $NS5$-branes
provides an interesting example of a time-dependent process
in string theory, which can be studied using a
Dirac-Born-Infeld effective field theory description. 

Some of the solutions of the DBI equations of motion describe the
D-brane falling towards the fivebranes; for non-zero angular
momentum density, the D-brane is spiralling towards the fivebranes.
These solutions exist when the energy and
angular momentum densities satisfy the inequality \boundangu, or
equivalently when the impact parameter satisfies the bound \condss.
In other solutions of the DBI equations of motion, the D-brane
comes in from infinity, is deflected by the fivebranes and escapes
back to infinity. We do not find solutions corresponding to a
D-brane in orbit around the fivebranes.

The real time dynamics of a D-brane near a stack of $NS5$-branes
exhibits a close connection to tachyon condensation on unstable
D-branes, and one can use results about the latter to better understand
the former. The effective potential for the radial coordinate $R$
goes exponentially to zero (in the natural variables) as the D-brane
approaches the fivebranes (\ie\ as $R\to 0$). As is well known from
the rolling tachyon problem, this means that the pressure goes
exponentially to zero as well. Thus, as the D-brane approaches
the fivebranes, it behaves like a pressureless fluid. This is very
similar to the late time behavior of unstable D-branes.

In the context of tachyon condensation, the pressureless fluid behavior
at late times is usually interpreted as a signal of the decay of the
unstable D-brane, and has been argued to provide a dual description
of the closed string radiation that the D-brane decays into. It is natural
to assume that a similar interpretation can be made here as well, with the
D-brane shedding its energy into modes living on the fivebranes as it
approaches them. We will see shortly that this interpretation is consistent
with spacetime expectations.

It is important to note that while the behavior of the D-brane near the
minimum of the potential provided by the $NS5$-branes is very similar
to that familiar from tachyon physics, its behavior near the top of the
potential is different. In particular, while the potential of an open string
tachyon has a quadratic maximum, corresponding to the original unstable
D-brane, at a finite value of $T$ (with a particular definition of the kinetic
term), in our case the maximum of the potential is at $T\to\infty$, in accord
with the interpretation of the potential at large $T$ as due to the gravitational
attraction between the D-brane and the fivebranes. Thus, features of the tachyon
analysis that depend on physics near the top of the potential are in general
different in our case.

A natural question is what happens to the RR charge of 
the $Dp$-brane if, as suggested by the DBI analysis, 
it loses all its energy as it approaches the fivebranes.
Consider, for example, the case $p=5$, in which a $D5$-brane of type IIB
string theory approaches a stack of $k$ $NS5$-branes, with all $k+1$
branes parallel. One expects the fivebranes to form a bound state, a
$(k,1)$ fivebrane, which is a member of the $SL(2,\Z)$ multiplet of
fivebranes of IIB string theory. For weak coupling, the energy density
of a $(k,1)$ fivebrane is (omitting constants)
\eqn\endensko{E_{k,1}\sim M_s^6\sqrt{\left(k\over g_s^2\right)^2+
{1\over g_s^2}}\simeq M_s^6\left({k\over g_s^2}+{1\over 2k}\right)~.}
We see that when a $D5$-brane binds to the $NS5$-branes, the binding
energy is almost 100\%.  As the $D5$-brane rolls in the fivebrane throat,
it must lose most of its energy density: out of $\tau_6\simeq M_s^6/g_s$,
only $\simeq M_s^6$ should be left, stored in a kind of ``remnant'' which
carries the RR charge of the original $D5$-brane. D-brane RR charges
in fivebrane backgrounds are discussed in \MaldacenaSS.

That is consistent with the picture we get from the DBI action.
The pressureless fluid that the D-brane approaches at late times
reflects the fact that most of the energy of the $D5$-brane
is radiated away as the D-brane rolls in the fivebrane throat.
It would be nice to show explicitly that the radiated amount is precisely
correct, but this is challenging even for the rolling tachyon solution.

For $p<5$ one can repeat the same analysis. It is convenient to
compactify the directions along the fivebranes on a five-torus,
and study the bound states of a $Dp$-brane wrapped around part
of the torus, with the fivebranes. The energetics of these bound
states is discussed in \SeibergZK, where it is shown that they
again have the property that the binding energy is almost 100\%,
as in \endensko, and that one can think of these bound states as
configurations carrying various fluxes in the gauge theory on the
$NS5$-branes. Thus, in all these cases, the rolling $Dp$-brane
must radiate most of its energy as it approaches the fivebranes,
in agreement with the DBI analysis.

Before concluding, we would like to briefly mention some additional
issues relevant to the discussion of this paper.

Our results are applicable to the case of two or more $NS5$-branes $(k\ge 2)$.
It is known that a single fivebrane does not develop the infinite throat \chs.
At the same time, we expect that an $NS5$-brane and a $D5$-brane should be
able to bind into a $(1,1)$ fivebrane, and in the process most of the energy
of the $D5$-brane should be radiated away, as in the discussion following
\endensko. It would be interesting to understand this process by analyzing the
dynamics of a $Dp$-brane in the background of a single $NS5$-brane.

Our discussion was based on the DBI effective action, and was thus approximate.
The DBI action is generally valid when the acceleration of the D-brane is small.
Thus, the late time behavior of the solution \asympsol, should be reliably
described by this action. Indeed, it is known from the study of rolling tachyons
that the corrections to \asympsol, which give rise to the acceleration, decay
exponentially with time. The full solution \solphi\ should be reliable for large
$k$, but may in general receive $1/k$ corrections. It could be that due to the
$N=2$ superconformal symmetry on the worldsheet, these corrections are absent,
but this has not been proven.

In \LukyanovNJ\ it was shown that a brane with the worldvolume \solphi\ (or, more
precisely, its Wick rotated Euclidean version) indeed exists and corresponds to an
exactly solvable boundary CFT in the bosonic string. In order to extend our analysis
of the solution \solphi\ to finite $k$, one needs to generalize the results of
\LukyanovNJ\ to the superconformal case. We hope to return to this problem elsewhere.

One can also ask whether there are large quantum (\ie\ $g_s$)
corrections to the classical open string analysis presented here. 
As we saw in section 3, the string coupling eventually becomes
large along trajectories such as \solradial, but if the energy
density of the D-brane is in the range \enconst, this happens
very late, and there is a large window in time, in which \philocal,
and thus string loop corrections, are small. Following the 
ideas of \refs{\SenXS,\SenIV}, one is led to the following
physical picture. The open string theory describing the rolling 
D-brane may provide a dual description of the closed string 
radiation produced as the D-brane moves towards the fivebranes. 
This radiation may be primarily produced during the long period
in which the string coupling \philocal\ is small, and thus it 
can be reliably described by perturbative string theory. 
The bound state of the D-brane and fivebranes that the system 
approaches at late times is a non-perturbative object, which 
may not be easy to study using (open or closed) weakly coupled 
string theory.

Some of the other questions raised by our results are:

\item{(1)} We found that there is a critical value of the 
angular momentum, $L=\sqrt{k}l_sE$, at which the dynamics 
of the D-brane changes considerably. Below that value, the 
D-brane always approaches at early or late times (or both) 
the vicinity of the fivebranes, $R=0$. For $L>\sqrt{k}l_sE$ 
the D-brane never approaches the fivebranes, and its radius 
of closest approach grows with $L$, \rrmin. As $L$ approaches 
the critical value, the exponential decay of the pressure 
with time is suppressed \solraddll, \pressll. It might be 
interesting to consider the dynamics near the critical angular
momentum.
\item{(2)} One of the reasons we did not find solutions corresponding
to D-branes orbiting around the $NS5$-branes was that at long
distances the gravitational potential of the fivebranes
goes like $-1/r^2$. If one compactifies one of the directions transverse
to the fivebranes on a circle, the potential will go like $-1/r$ at long
distances. It would be interesting to investigate the D-brane trajectories
in that case, and in particular to look for stable orbits (see
\BurgessQV\ for a discussion of such orbits for other D-brane systems).
\item{(3)} There are generalizations of the CHS geometry \chs\ that
describe other singularities in string theory, such as generalized
conifold singularities of Calabi-Yau surfaces (see \eg\ \GiveonZM).  
It would be interesting to repeat the analysis of this paper
for such geometries, and for their regularized versions, which
are relevant for Little String Theory \refs{\AharonyUB,\GiveonPX}.  
\item{(4)} Cosmological applications of D-brane motion in warped
throats in string theory have been extensively studied recently
(see \eg\ 
\refs{\KehagiasVR,\BurgessTZ,\SilversteinHF,\DeWolfeQX,\AlishahihaEH,\BuchelQG} 
for some recent discussions and additional references). 
It would be interesting to investigate
what types of cosmologies arise from D-branes rolling down
the throats of $NS5$-branes, and how they are related to
tachyon driven cosmologies of the sort discussed \eg\ in
\refs{\GibbonsMD,\ShiuXP,\SenMV} and references therein.
In particular, we saw that naively extrapolating the trajectory
of the $Dp$-brane through the strong coupling region, an observer
on the D-brane sees a cyclic cosmology \soltau, in which 
the D-brane oscillates about the fivebranes. It would be 
interesting to study the corrections to this picture in the
full quantum theory.
\item{(5)} The D-brane of \LukyanovNJ\ \euclbrane\ and its
Minkowski continuation \solphi\ can be studied in two 
dimensional string theory, which has a matrix model dual. 
What is the interpretation of this D-brane and its dynamics there?

\bigskip
\noindent{\bf Acknowledgements:}
I am grateful to O. Aharony, J. Harvey, F. Larsen, E. Martinec,
V. Niarchos, G. Rajesh, D. Sahakyan and A. B. Zamolodchikov for 
useful discussions. I also thank the NHETC at Rutgers University
for hospitality. This work was supported in part by DOE grant 
\#DE-FG02-90ER40560.

\listrefs
\end